\documentstyle[12pt,aps,epsf]{revtex}
\begin{document}
\tighten
\draft

\title{ Superextendons with a modified measure                           
\protect\\  } \author{E.I. Guendelman                                         
\\{\it Physics Department, Ben-Gurion University, Beer-Sheva                  
84105, Israel}}                                                               
                                                                              
\maketitle                                                                    
\bigskip                                                                      
\begin{abstract}                                                              
For superstrings, the consequences of replacing the measure of 
integration $\sqrt{-\gamma}d^2 x$ in the Polyakov's action by $\Phi d^2 x$
where $\Phi$ is a  density built out of degrees of freedom 
independent of the metric $\gamma_{ab}$ defined in the string are 
studied. As in Siegel reformulation of the Green Schwarz formalism the  
Wess-Zumino term is the square of supersymmetric currents. As opposed
to the Siegel case, the compensating fields needed for this do not
enter into the action just as in a total derivative. They instead play
a crucial role to make up a consistent dynamics.
The string tension appears as an integration constant of the 
equations of motion. The generalization to higher dimensional extended
objects is also studied using in this case the Bergshoeff and Sezgin
formalism with the associated additional fields, which again are 
dynamically relevant unlike the standard formulation. Also unlike the 
standard formulation, there is no need of a cosmological term on the 
world brane.     
                                   
\end{abstract}

\section{Introduction}
There are reasons to consider changing the way we think and formulate
generally covariant theories. If these are generally covariant
theories of gravity, one of these reasons is the cosmological constant 
problem\cite{Wein}, which is a consequence of the fact that in gravitational theory,
as usually formulated, the origin of the energy density is important.

This is very much related to the fact that the action for generally
covariant theories, which is usually writen as

\begin{equation}                      
S = \int d^{d}x \sqrt{-\gamma}  L
\end{equation}  

where $\gamma$ is the determinant of the metric and $L$ is a scalar,
is not invariant under the shift $ L \rightarrow L+ const.$ .

If in (1) we were to change the measure of integration $d^{d}x \sqrt{-\gamma}$
by $d^{d}x \Phi $, where $\Phi $ is a total derivative, then the shift
$ L \rightarrow L+ const.$  will indeed be a symmetry.

This possibility was studied in the context of gravitational theories 
which can handle the cosmological constant problem \cite{modmes} and as 
a tool for the construction of new types of scale invariant theories 
consistent with non trivial masses and potentials which are of the 
form required by inflation \cite{ind-mea} ,\cite{decl}, \cite{cosco},
 \cite{gold} . In the models of
Refs. \cite{ind-mea} ,\cite{decl}, \cite{cosco} , \cite{gold} no fundamental 
dimensionfull parameter really appears in the fundamental lagrangian (one
can indeed introduce such parameters, but they can be reabsorbed for
example by a rescaling of the fields that define the measure $\Phi$).

It is very interesting that the issues raised above in the case of 
gravitational theories have their analogs in string and brane theories,
even before we attempt using these theories as theories of gravity,
that is we are talking here of string or brane world sheet analogs of the 
issues raised above.

To begin with, string and brane theory \cite{Yuval} have appeared as candidates for unifying all interactions 
of nature. One aspect of string and brane theories seems to many not quite appealing
however: this is the introduction from the begining of a fundamental scale,
the string or brane tension.
The idea that the fundamental theory of nature, whatever that may be, 
should not contain any fundamental scale has attracted a lot of attention.
According to this point of view, whatever scale appears in nature, must not
appear in the fundamental lagrangian of physics. Rather, the appearence 
of these scales must be spontaneous, for example due to boundary conditions
in a classical context or a process of dimensional transmutation to give an
example of such effect in the context of quantum field theory. Such issue,
that is the spontaneous appearence of a scale in scale invariant theories
which use a modified measure was addressed (in a four dimensional context)
and satisfactory solved in Refs. \cite{ind-mea} ,\cite{decl}, \cite{cosco} ,
 \cite{gold}.

Interestingly enough, the cosmological constant of gravitational theories
has its  analog in brane theory. It is well known that the generalization of
the Polyakov formalism to branes must incorporate an explicit world brane
cosmological term, unlike the string case, where such term is forced to vanish.
A definite asymmetry between string and higher branes gets established this way.              

Here we will see what we obtain from string and brane theories
with a modified measure. As we will see, string theories or more 
generally brane theories without a fundamental scale are possible if the 
extended objects do not have boundaries (i.e., they are closed). Also for
higher dimensional branes no explicit world brane cosmological constant 
needs to be included, therefore restoring the symmetry between strings 
and branes. How the 'brane cosmological constant problem' is related to the
cosmological constant problem of the gravitational low energy theory is 
not known, but that connection may very well exist. 

The situation for bosonic strings and branes with a modified measure was already sudied 
in a previous work \cite{strm}. Here we review this and proceed then to 
generalize this to the supersymmetric case. 

\section{Bosonic string and brane theories with a modified measure}
In this section we review the previous work \cite{strm} on bosonic
extendons with a modified measure before going into the supersymmetric
case.
The Polyakov action for the bosonic string is \cite{Poly}

 \begin{equation}
 S_{P}[X,\gamma_{ab}] =   -T \int d\tau d\sigma \sqrt{-\gamma}
\gamma^{ab} \partial_{a} X^{\mu}\partial_{b} X^{\nu} g_{\mu \nu}                              
 \end{equation}
                                                
Here $\gamma_{ab}$ is the metric defined in the $1+1$ world sheet 
of the string and $ \gamma = det(\gamma_{ab})$. $g_{\mu \nu}$ is the 
metric of the embedding space. $T$ is here the string tension, a 
dimensionfull quantity introduced into the theory, which defines a
scale.

We recognize the measure of integration
$d\tau d\sigma \sqrt{-\gamma}$ and as we anticipated before, we want 
to replace this measure of integration by another one which does 
not depend on $\gamma_{ab}$ .  

If we introduce two scalars (both from the point of view of the $1+1$
world sheet of the string and from the embedding $D$-dimensional universe)
$\varphi_{i}$, $i=1,2$, we can construct the world sheet density

\begin{equation}                                                              
\Phi =  \varepsilon^{ab}  \varepsilon_{ij}                   
\partial_{a} \varphi_{i} \partial_{b} \varphi_{j}                                      
\end{equation}                                                                
                                                                              
where $\varepsilon^{ab}$ is given by $\varepsilon^{01} = 
-\varepsilon^{10} =1$, $\varepsilon^{00} =                      
\varepsilon^{11} = 0 $ and $\varepsilon_{ij} $ is defined by
$\varepsilon_{12} = -\varepsilon_{21} = 1$, $\varepsilon_{11} =
\varepsilon_{22} =0 $. 

It is interesting to notice that $d\tau d\sigma \Phi = 
2 d \varphi_{1} d \varphi_{2}$, that is the measure of integration
$d\tau d\sigma \Phi $ corresponds to integrating in the space 
of the scalar fields $\varphi_{1}, \varphi_{2}$.

We proceed now with the construction of an action that uses 
$d\tau d\sigma \Phi$ instead of $d\tau d\sigma \sqrt{-\gamma}$. 
When considering the types of actions we can have under these
circumtances, the first one that comes to mind ( a straightforward
generalization of the Polyakov action) is

\begin{equation}
S_{1} = - \int d\tau d\sigma \Phi
\gamma^{ab} \partial_{a} X^{\mu}\partial_{b} X^{\nu} g_{\mu \nu} 
\end{equation} 
          
Notice that multiplying $S_{1}$ by a constant, before boundary or initial
conditions are specified is a meaningless operation, since such a constant 
can be absorbed in a redefinition of the measure fields $\varphi_{1}, 
\varphi_{2}$ that appear in $\Phi$.

The form (4) is however not a satisfactory action, because the variation of 
$S_{1}$ with respect to $\gamma^{ab}$ leads to the rather strong condition

\begin{equation}
\Phi \partial_{a} X^{\mu}\partial_{b} X^{\nu} g_{\mu \nu} = 0
\end{equation}            

If $\Phi \neq 0 $, it means that 
$ \partial_{a} X^{\mu}\partial_{b} X^{\nu} g_{\mu \nu} = 0$, which means
that the metric induced on the string vanishes, clearly not an acceptable
dynamics. Alternatively, if $\Phi=0$, no further information is available, 
also a not desirable situation.
                                                                                                                                                                
To make further progress, it is important to notice that terms that when 
considered as contributions to $L$ in  
                               
\begin{equation}                                                              
S = \int d\tau d\sigma \sqrt{-\gamma}  L                   
\end{equation}                                                                
                                        
which do not contribute to the equations of motion, i.e., such that 
$\sqrt{-\gamma}  L$ is a total derivative, may contribute when we 
consider the same $L$, but in a contribution to the action of the form

\begin{equation}        
S = \int d\tau d\sigma \Phi L
\end{equation} 

This is so because if $\sqrt{-\gamma}  L$ is a total divergence, $ \Phi L$ 
in general is not.

This fact is indeed crucial and if we consider an abelian gauge field 
$A_{a}$ defined in the world sheet of the string, in addition to the 
measure fields $\varphi_{1},         
\varphi_{2}$ that appear in $\Phi$, the metric $\gamma^{ab}$ and the
string coordinates $ X^{\mu}$, we can then construct the non trivial
contribution to the action of the form

\begin{equation}
 S_{gauge} = \int d\tau d\sigma \Phi 
\frac {\varepsilon^{ab}}{\sqrt{-\gamma}} F_{ab}
\end{equation}                                       
 where 
\begin{equation} 
F_{ab} = \partial_{a} A_{b} - \partial_{b} A_{a}
\end{equation}

So that the total action to be considered is now
\begin{equation}   
S = S_{1} +  S_{gauge} 
\end{equation} 

 with $S_{1}$ defined as in eq. 4 and $S_{gauge}$ defined by eqs.8 and 9.

The action (10) is invariant under a set of diffeomorphisms in the space
of the measure fields combined with a conformal transformation of the metric
$\gamma_{ab}$,
\begin{equation} 
\varphi_{i} \rightarrow \varphi_{i}^{'} = \varphi_{i}^{'} (\varphi_{j})
\end{equation}

So that,

\begin{equation}
\Phi \rightarrow \Phi^{'} = J \Phi 
\end{equation} 
where J is the jacobian of the transformation (11)
and 
\begin{equation}
\gamma_{ab} \rightarrow \gamma^{'}_{ab} = J \gamma_{ab}
\end{equation} 
        
The combination $\frac {\varepsilon^{ab}}{\sqrt{-\gamma}} F_{ab} $ is a 
genuine scalar. In two dimensions is proportional to $\sqrt{ F_{ab} F^{ab}}$.

Working with (10), we get the following equations of motion:
From the variation of the action with respect to $\varphi_{j}$

\begin{equation}
\varepsilon^{ab} \partial_{b} \varphi_{j} \partial_{a} (
-\gamma^{cd} \partial_{c} X^{\mu}\partial_{d} X^{\nu} g_{\mu \nu} +
\frac {\varepsilon^{cd}}{\sqrt{-\gamma}} F_{cd} ) = 0
\end{equation}

If $det (\varepsilon^{ab} \partial_{b} \varphi_{j}) \neq 0$, which
means $\Phi \neq 0$, then we must have that all the derivatives of 
the quantity inside the parenthesis in eq.14 must vanish, that is, 
such a quantity must equal a constant which will be determined later,
but which we will call $M$ in the mean time,   

\begin{equation} 
-\gamma^{cd} \partial_{c} X^{\mu}\partial_{d} X^{\nu} g_{\mu \nu} +
\frac {\varepsilon^{cd}}{\sqrt{-\gamma}} F_{cd} = M
\end{equation}  

The equation of motion of the gauge field $A_{a}$, tells us about 
how the string tension appears as an integration constant. 
Indeed this equation is

\begin{equation} 
\varepsilon^{ab} \partial_{b} (\frac {\Phi}{\sqrt{-\gamma}}) = 0
\end{equation}

which can be integrated to give

\begin{equation}
\Phi = c \sqrt{-\gamma}
\end{equation}

Notice that (17) is perfectly consistent with the conformal symmetry
 (11), (12) and (13). Equation 15 on the other hand is consistent with 
such a symmetry only if $M = 0$. Indeed, we will check that the equations 
of motion indeed  imply that $M = 0$. In the case of higher dimensional 
branes, the equations of motion will imply that  $M$ is non vanishing.

By calculating the Hamiltonian, after dropping boundary terms (this is
totally justified in the case of closed strings) and (only at the end of
the process) using eq.17,  we find that $c$ equals the string 
tension.

Now let us turn our attention to the equation of motion derived from the
variation of (10) with respect to $\gamma^{ab}$. We get then,

\begin{equation}
 - \Phi (\partial_{a} X^{\mu}\partial_{b} X^{\nu} g_{\mu \nu}
- \frac {1}{2} \gamma_{ab} \frac {\varepsilon^{cd}}{\sqrt{-\gamma}} F_{cd}) = 0  
\end{equation}  

From the constraint (15), we can solve 
$\frac {\varepsilon^{cd}}{\sqrt{-\gamma}} F_{cd} $ and insert back into 
(18), obtaining then (if $\Phi \neq 0$)

\begin{equation}
\partial_{a} X^{\mu}\partial_{b} X^{\nu} g_{\mu \nu} -
 \frac {1}{2} \gamma_{ab} 
\gamma^{cd} \partial_{c} X^{\mu}\partial_{d} X^{\nu} g_{\mu \nu}
-\frac {1}{2} \gamma_{ab} M = 0
\end{equation}

Multiplying the above equation by  $\gamma^{ab}$ and summing over $a, b$,
we get that $M = 0$, that is the equations are exactly those of the Polyakov action. 
After eq.16 is used, the eq. obtained from the variation of $X^{\mu}$ is
seen to be exactly the same as the obtained from the Polyakov action as well.

Let us now consider a $d+1$ extended object, described (generalizing the 
action (9)),

\begin{equation} 
S = S_{d} + S_{d-gauge}
\end{equation}  

where 
\begin{equation}
 S_{d} = -\int d^{d+1} x \Phi                                             
\gamma^{ab} \partial_{a} X^{\mu}\partial_{b} X^{\nu} g_{\mu \nu}
\end{equation}
and 

\begin{equation}
S_{d-gauge} =  \int d^{d+1} x \Phi
\frac {\varepsilon^{a_{1}a_{2}...a_{d+1}}}{\sqrt{-\gamma}} 
\partial_{[a_{1}}A_{a_{2}...a_{d+1}]}
\end{equation} 

and
\begin{equation}
\Phi = \varepsilon^{a_{1}a_{2}...a_{d+1}}\varepsilon_{j_{1}j_{2}...j_{d+1}}
\partial_{a_{1}} \varphi_{j_{1}}....\partial_{a_{d+1}} \varphi_{j_{d+1}}
\end{equation} 

This model does not have a symmetry which involves an arbitrary 
diffeomorphism in the space of the measure fields coupled with a conformal
transformation of the metric, except if $ d=1 $ (eqs. (11), (12) and (13)). 
For $d \neq 1$, there is 
still a global
scaling symmetry where the metric transforms as ($\theta$ being a constant),
                  
\begin{equation}                                                              
\gamma_{ab}  \rightarrow   e^{\theta} \gamma_{ab}                              
\end{equation}                                                                

the $\varphi_{j}$ are transformed according to                                 

\begin{equation}
\varphi_{j}   \rightarrow   \lambda_{j} \varphi_{j}                         
\end{equation} 

(no sum on $j$) which means                                                     
$\Phi \rightarrow \biggl(\prod_{j} {\lambda}_{j}\biggr) \Phi \equiv \lambda   
\Phi $

Finally, we must demand that $\lambda = e^{\theta} $  and that the transformation
of $A_{a_{2}...a_{d+1}}$ be defined as

\begin{equation} 
A_{a_{2}...a_{d+1}} \rightarrow \lambda ^{\frac {d-1}{2}}A_{a_{2}...a_{d+1}}
\end{equation}  

Then we have a symmetry. Also no scale is introduced into the theory from 
the beginning. This is apparent from the fact that any constants multiplying
the separate contributions to the action  (21) or (22) is meaningless if no 
boundary or initial conditions are specified, because then such factors can 
be absorbed  by a redefinition of the measure fields and of the gauge fields.
Notice that the existence of a symmetry alone is not enough to guarantee that
no fundamental scale appears in the action. For example string theory, 
as usually formulated has conformal symmetry, but the string tension is 
still a fundamental scale in the theory.

Another interesting symmetry of the action (up to the integral of a total 
divergence) consists of the infinite dimensional set of transformations                                        
$\varphi_{j} \rightarrow \varphi_{j} + f_{j} (L)$, 
where $f_{j} (L)$ are arbitrary functions of
 
\begin{equation}
L = - \gamma^{cd} \partial_{c} X^{\mu}\partial_{d} X^{\nu} g_{\mu \nu} +              
 \frac {\varepsilon^{a_{1}a_{2}...a_{d+1}}}{\sqrt{-\gamma}}                   
\partial_{[a_{1}}A_{a_{2}...a_{d+1}]}                     
\end{equation} 

This symmetry does depend on the explicit form of the lagrangian density
$L$ , but only the fact that  $L$ is $\varphi_{a}$ independent.

Now we go through the same steps we went through in the case of the string.
The variation with respect to the measure field $\varphi_{j}$ gives 

\begin{equation}
K^{a}_{j} \partial_{a} (                      
-\gamma^{cd} \partial_{c} X^{\mu}\partial_{d} X^{\nu} g_{\mu \nu}
+ \frac {\varepsilon^{a_{1}a_{2}...a_{d+1}}}{\sqrt{-\gamma}}                    
\partial_{[a_{1}}A_{a_{2}...a_{d+1}]}) = 0
\end{equation}

where 
\begin{equation}                                                              
K^{a}_{j} = \varepsilon^{a a_{2}...a_{d+1}}\varepsilon_{j j_{2}...j_{d+1}}   
\partial_{a_{2}} \varphi_{j_{2}}....\partial_{a_{d+1}} \varphi_{j_{d+1}}      
\end{equation}                 

Since $det(K^{a}_{j})= \frac {(d+1)^{-(d+1)}}{(d+1)!}\Phi^{d}$, it therefore 
follows that for $\Phi \neq 0$, $det(K^{a}_{j})\neq 0$ and

\begin{equation}  
-\gamma^{cd} \partial_{c} X^{\mu}\partial_{d} X^{\nu} g_{\mu \nu} 
+ \frac {\varepsilon^{a_{1}a_{2}...a_{d+1}}}{\sqrt{-\gamma}}
\partial_{[a_{1}}A_{a_{2}...a_{d+1}]} = M 
\end{equation}

where $M$ is some constant of integration. If $d \neq 1$ then $M \neq 0$ as we
will see. Furthermore, under a scale transformation (24), (25) and (26),
$M$ does change from one constant value to another.

The variation with respect to the gauge field $A_{a_{2}...a_{d+1}}$ leads to
the equation 

\begin{equation}
\varepsilon^{a_{1}a_{2}...a_{d+1}}\partial_{a_{1}}
\frac {\Phi}{\sqrt{-\gamma}} = 0 
\end{equation}  

which means

\begin{equation}                                                              
\Phi = c \sqrt{-\gamma}                                                       
\end{equation}                                                                

once again. As in the case of the string the brane tension has been generated
spontaneously instead of appearing as a parameter of the fundamental 
lagrangian. Again a simple calculation of the Hamiltonian and using after this
the above equation, we obtain that $c$ is proportional to the brane tension.

The variation of the action with respect to $\gamma^{ab}$ leads to

\begin{equation}
- \Phi (\partial_{a} X^{\mu}\partial_{b} X^{\nu} g_{\mu \nu}               
- \frac {1}{2} \gamma_{ab} 
\frac {\varepsilon^{a_{1}a_{2}...a_{d+1}}}{\sqrt{-\gamma}} 
\partial_{[a_{1}}A_{a_{2}...a_{d+1}]}) = 0 
\end{equation}  

We can now solve for 
$\frac {\varepsilon^{a_{1}a_{2}...a_{d+1}}}{\sqrt{-\gamma}}
\partial_{[a_{1}}A_{a_{2}...a_{d+1}]}$ from equation (30) and then reinsert 
in the above equation, obtaining then, 

\begin{equation} 
\partial_{a} X^{\mu}\partial_{b} X^{\nu} g_{\mu \nu} =
\frac {1}{2} \gamma_{ab} 
(\gamma^{cd}\partial_{c} X^{\mu}\partial_{d} X^{\nu} g_{\mu \nu} + M )
\end{equation} 

This is the same equation that we would have obtained from a Polyakov type
action augmented by a cosmological term.

As in the case of the string, $M$ can be found by contracting both sides of
the equation. For $d \neq 1 $, $M$ is non zero and equal to
\begin{equation}
M = 
\frac{\gamma^{cd}\partial_{c} X^{\mu}\partial_{d} X^{\nu} g_{\mu \nu} (1-d)}{1+d}
\end{equation}

We can also solve for 
$\gamma^{cd}\partial_{c} X^{\mu}\partial_{d} X^{\nu} g_{\mu \nu}$ in terms of 
$M$ from (35) and insert in the right hand side of (34), obtaining,

\begin{equation}
\gamma_{ab} = \frac {1-d}{M}
\partial_{a} X^{\mu}\partial_{b} X^{\nu} g_{\mu \nu}
\end{equation}

Which means that $\gamma_{ab}$ is up to the constant 
factor $\frac {1-d}{M}$ equal
to the induced metric. Since there is the scale invariance  (24), (25) 
and (26) an overall constant factor in the evolution of $\gamma_{ab}$ cannot be 
determined. The same scale invariance means however that there is a field
redefinition which does not affect any parameter of the lagrangian and
which allows us to set $\gamma_{ab}$ equal to the induced metric (at least
if we start from any negative value of $M$), that is,

\begin{equation}
\gamma_{ab} = \partial_{a} X^{\mu}\partial_{b} X^{\nu} g_{\mu \nu} 
\end{equation}

In such case $M$ is consistently given (inserting (37) into (36) or (35)),

\begin{equation} 
M = 1-d
\end{equation} 

Notice that in contrast with the standard approach for Polyakov type actions 
in the case of higher dimensional branes \cite{lambda}, here we do not have to 
fine tune a 
parameter of the lagrangian the brane "cosmological constant", so as to 
force that (37) be satisfied. Here instead, it is an integration constant,
that appears from an action without an original cosmological term, which can be
set to the value given by eq. (38) by means of a scale transformation. Such 
choice ensures then that (38) is satisfied (and therefore (37)).
Furthermore, it appears that this 
treatment is more appealing if one thinks of all branes on similar footing, 
since in the approach of this paper they can all be described by a similar 
looking lagrangian, unlike in the usual aproach which discriminates in a 
radical way between strings, these having no cosmological constant 
associated to them, and the higher dimensional branes, which require a
fine tuned cosmological constant.

If we do not make the choice (38), the constant $c$ is not directly the 
brane tension, which is instead $c(\frac{1-d}{M})^{\frac{d-1}{2}}$, 
which can be checked is scale invariant combination, since under a scale
transformation $M \rightarrow e^{\theta} M $ and
$c \rightarrow e^{\theta \frac{(d-1)}{2}}c$. If the choice 
$M = 1-d $ is made the brane tension is simply $c$. The principle remains 
the same as in the case of the string and the constant $c$ 
is still responsible for the 
spontaneous generation of a brane tension.

One should notice that other authors have also constructed 
actions for branes that do not contain a brane-cosmological term \cite{Dolan}.
Such formulations depend, unlike what has been developed here, on the 
dimensionality, in particular whether this is even or odd, so that it is 
clear that those formulations do not have much relation with what has 
been done here. Yet other approaches \cite{Barc}
to an action without a brane cosmological involve lagrangians with non linear
dependence on the invariant 
$\gamma^{cd}\partial_{c} X^{\alpha}\partial_{d} X^{\beta} g_{\alpha \beta}$,
also a rather different path to the one followed here. For an interesting
analysis of  different possible Lagrangians for extendons 
see \cite{Eiz}.

An approach that has some common features to the one developed here is
that of Refs. \cite{Berg} and \cite{Anso} where also the tension of the brane is found 
as an integration constant. Here also gauge fields are introduced, but they
appear in a quadratic form rather than in a linear form. Also  scale
invariance is discussed in \cite{Berg}, but it is a target space scale 
invariance since no metric defined in the brane is studied there, 
i.e. no connection to a 
Polyakov type action, which is known to be more useful in the
quantum theory, is made. For the case of superstrings and superbranes we
will follow a procedure that keeps the basic structure found in the bosonic
case, except that the gauge fields introduced here in order to obtain a 
consistent dynamics turn out to be composites of other fields, a rather
different approach to that of Ref. \cite{Berg}. The linearity of the 
lagrangian on the (in this case composite) gauge fields will be mantained,
also unlike Ref. \cite{Berg} (Ref. \cite{Anso} does not discuss the 
super symmetric case). 

\section{Superstrings with a modified measure}

The general structure that we have found for the bosonic strings and 
branes suggest what is the way to follow in the case of superstrings
and supermembranes.

In fact, the additional term with the gauge fields defined by eqs.(8), (9),
being associated with the alternating symbol in two dimensions, appears very
much related to the Wess Zumino term in the Green Shwarz formulation of the
superstring \cite{GS} .

It is important to notice that in the Green Schwarz formulation, the 
Wess Zumino term is not invariant under supersymmetry, but only invariant
up to a total divergence. And since we have already discussed, in our
formulation total derivatives have to be handled with care, since 
if $\sqrt{-\gamma}  L$ is a total divergence, $ \Phi L$ 
in general is not.

Under these circumtances, Siegel \cite{Siegel} reformulation of the
Green Schwarz superstring, where the Wess Zumino term is manifestly
supersymmetric becomes of special interest from our point of view.

In Siegel \cite{Siegel} reformulation, the action of the superstring in a flat 
embedding spacetime (with metric $\eta_{\mu \nu}$) is
written as   

\begin{equation}
S = T \int L \sqrt{-\gamma}  d^{2}x
\end{equation}   

where the scalar $L$ is given by

\begin{equation}
L = -\frac{1}{2} \gamma^{ab} J^{\mu}_{a} J^{\nu}_{b} \eta_{\mu \nu} 
-i\frac {\varepsilon^{a b}}{\sqrt{-\gamma}} J^{\alpha}_{a} J_{\alpha b} 
\end{equation} 
where the currents $J^{\mu}_{a}$, $J^{\alpha}_{a}$ and 
$J_{\alpha b}$ are defined as
\begin{equation} 
J^{\alpha}_{a} = \partial_{a} \theta^{\alpha}, 
\end{equation} 
\begin{equation} 
J^{\mu}_{a} = \partial_{a} X^{\mu} -
 i \partial_{a} \theta^{\alpha}\Gamma^{\mu}_{\alpha \beta} \theta^{\beta},
\end{equation}
\begin{equation} 
J_{\alpha a} = \partial_{a} \phi_{\alpha} -
2i(\partial_{a} X^{\mu})\Gamma_{\mu \alpha \beta} \theta^{\beta}
- \frac{2}{3} (\partial_{a} \theta^{\beta}) \Gamma^{\mu}_{ \beta \delta} 
\theta^{\delta}\Gamma_{\mu \alpha \epsilon} \theta^{\epsilon}
\end{equation} 

Then, there is the following symmetry of the lagrangian (exact, not up
to total divergence)
\begin{equation}
\delta \theta^{\alpha} = \epsilon^{\alpha}, 
\delta X^{\mu} =
 -i \epsilon^{\alpha}\Gamma^{\mu}_{ \alpha \beta}\theta^{\beta}, 
\end{equation}
and
\begin{equation} 
\delta \phi_{\alpha} = 2i \epsilon^{\beta} \Gamma _{\mu \alpha \beta} X^{\mu}
+ \frac{2}{3} (\epsilon^{\beta} \Gamma ^{\mu}_ {\beta \epsilon} 
\theta ^{\epsilon}) \Gamma _{\mu \alpha \kappa} \theta^{\kappa}
\end{equation}

If the Dirac $\Gamma^{\mu}_{ \alpha \beta}$ matrices satisfy a condition which
requires the target space dimensionality to be 3, 4, 6 or 10.
In this formalism the fields $\phi_{\alpha}$ are not determined at all
by the equations of motion, since their dependence enters only as a total
divergence.

The situation changes if in eq. (39) , $T\sqrt{-\gamma}  d^{2}x  \rightarrow
\Phi d^{2}x $ and we now consider 

\begin{equation} 
S = \int L \Phi d^{2}x
\end{equation}  
with  $L$ still given by eqs. (40), (41), (42) and (43).

It is now crucial to recognize that the abelian gauge field defined in the
world sheet of the string, which was introduced in order to obtain sensible
equations of motion in the case of the bosonic string, appears here induced 
by the additional fields introduced by Siegel.  
 
The identification of the abelian gauge field proceeds according to the
equation
  
 \begin{equation}
  -i \varepsilon^{a b} \partial_{a} \theta^{\alpha} 
 \partial_{b} \phi_{\alpha} \equiv   \varepsilon^{a b} \partial_{a} A_{b}
\end{equation}    

which can be indeed solved by the composite gauge field construction

\begin{equation} 
A_{b} \equiv  -i \theta^{\alpha}  \partial_{b} \phi_{\alpha}
\end{equation} 

Such composite gauge field construction is indeed very closely 
related to the ones studied by  E.I.Guendelman, E.Nissimov and 
S.Pacheva and also by C.Castro in Ref. \cite{modmes}.

One can then see that if no singularities or degenerate situations
are present, then all of the equations, except one, are the same as
those obtained in the standard Siegel formulation of the Green Schwarz
superstring \cite{Siegel}.

The difference is due to the fact that the abelian gauge field $A_{b}$, 
and therefore the $\phi_{\alpha}$ fields play a dynamical role, unlike the
case of the Siegel formalism. Unlike  the Siegel formalism, there is an 
equation that tells us something about the $\phi_{\alpha}$ fields. This
is the equation obtained from the variation of the measure fields.

\begin{equation}
\varepsilon^{ab} \partial_{b} \varphi_{j} \partial_{a} L = 0 
\end{equation} 
   
which means, if $\Phi \neq 0 $ that 

\begin{equation} 
L = M = constant
\end{equation}

The variation with respect to $\phi_{\alpha}$ gives the equation

\begin{equation}
\varepsilon^{ab} \partial_{a} \theta^{\alpha} 
\partial_{b} (\frac {\Phi}{\sqrt{-\gamma}}) = 0
\end{equation}

If we have a non degenerate situation, that is for enough linearly 
independent non vanishing components of $\partial_{a} \theta^{\alpha}$,
it follows that,
\begin{equation}
\frac {\Phi}{\sqrt{-\gamma}} = c = constant.
\end{equation} 

and as in the bosonic case, the integration constant 
 $c$ is the string tension.

Following the steps of section $2$, we can once again find, by combining
(50) and the equation obtained from the variation with respect to the
world sheet metric that
 
\begin{equation}   
M = 0
\end{equation}   

As anticipated, once (52) is used, all the resulting equations are exactly 
those in Ref. \cite{Siegel} , except for (50) with $M = 0$, i.e. the vanishing on the 
mass shell of the lagrangian. Such condition imposses a constraint on the 
$\phi_{\alpha}$ fields, which in \cite{Siegel} 
are totally undetermined.

The interesting role of the new fields $\phi_{\alpha}$ in obtaining a perfect
balance, so as to ensure that the lagrangian is exactly zero, may very well
be connected to a resolution of the cosmological constant is the effective
low energy gravitational theory. Recall that the ideas of using a modified
measure were motivated in the first place in this context \cite{modmes}.

\section{Superbranes with a modified Measure and without a Cosmological Term}
For higher dimensional superbranes Bergshoeff and Sezgin \cite{Superb} have 
generalized the auxiliary field formalism of Siegel.

As we saw in the bosonic case, the new feature that appears when 
considering higher dimensional branes, instead of strings is that in the
usual Polyakov type formalism, a wold brane cosmological term must be
included, but when the modified measure is used, no explicit cosmological
term is required. Instead, when the equations of motion are considered,
we are forced to consider a non vanishing value of the constant of
integration $M$. These features are mantained when we formulate the 
supermembrane generalization of the above.

We begin our discussion of the higher dimensional branes with the 
consideration of the $2+1$ dimensional brane, which will be treated 
in some detail. Once this is understood, the higher dimensional cases 
follow more or less in a similar fashion, provided the results
of Ref. \cite{Superb} are properly applied.

Once again, as in the superstring case, we want to write the lagrangian as
the sum of products of invariant supercurrents. For this to be achieved,
we need to introduce, in the case of the $2+1$ brane, the additional fields
$\phi_{\mu \nu}$ (field with two target space indices), $\phi_{\mu \alpha}$
(field with one target space index and one spinor index) and also
$\phi_{\alpha \beta}$ (field with two spinor indices), in addition to the
original $\theta^{\alpha}$ and $X^{\mu}$ fields of the brane.

Then we are in a position to define the currents
(where an abreviated notation is used in what follows, like 
$\overline {\theta} \Gamma^{\mu} \partial_{b}\theta$ being a short cut for 
$\theta^{\alpha}  \Gamma^{\mu}_{\alpha \beta} \partial_{b}\theta^{\beta}$, 
etc. . Also in this section we follow normalizations of the $\theta$ fields 
and other conventions  
of Ref.\cite{Superb} rather than those of Ref. \cite{Siegel})
\begin{equation} 
L^{\alpha}_{a} = \partial_{a} \theta^{\alpha},
\end{equation} 
\begin{equation}
L^{\mu}_{a} = \partial_{a} X^{\mu} + \frac{1}{2}
\overline {\theta}\Gamma^{\mu} \partial_{a} \theta,
\end{equation} 
\begin{equation}
L_{a \mu \nu} = \partial_{a} \phi_{\mu \nu} +
\frac{1}{2} \overline {\theta} \Gamma _{\mu \nu} \partial_{a} \theta,
\end{equation} 
\begin{eqnarray} 
L_{a \mu \alpha} = \partial_{a} \phi_{\mu \alpha} + 
\partial_{a} \phi_{\mu \nu} (\Gamma^{\nu} \theta)_{\alpha} + 
 \partial_{a}X^{\nu} (\Gamma_{\mu \nu} \theta)_{\alpha}
+\frac {1}{6} (\Gamma_{\mu \nu} \theta)_{\alpha} \overline{\theta} 
\Gamma^{\nu} \partial_{a} \theta \nonumber  \\ 
+\frac {1}{6}(\Gamma^{\nu}\theta)_{\alpha} \overline{\theta}\Gamma_{\mu \nu}
\partial_{a} \theta,
\end{eqnarray}
\begin{eqnarray}
L_{a \alpha \beta} = \partial_{a} \phi_{ \alpha \beta} - 
\frac {1}{2} X^{\mu} \partial_{a} 
\phi_{\mu \nu}(\Gamma^{\nu})_{ \alpha \beta}
+ \partial_{a} \phi_{\mu \nu} (\Gamma^{\mu}\theta)_{(\alpha}
(\Gamma^{\nu}\theta)_{\beta)} + \frac{1}{4}(\overline{\theta}\partial_{a}
\phi_{\mu})(\Gamma^{\mu})_{\alpha \beta} 
\nonumber  \\
+ 2 (\Gamma^{\mu}\theta)_{(\alpha} \partial_{a} \phi_{\mu \beta)}    
-\frac{1}{2}X^{\mu} \partial_{a} X^{\nu} (\Gamma_{\mu \nu})_{\alpha \beta}
-(\Gamma^{\nu}\theta)_{(\alpha}  (\Gamma_{\mu \nu}\theta)_{\beta)} 
\partial_{a}X^{\mu}          \nonumber     \\        
-\frac{1}{12}(\Gamma_{\nu}\theta)_{(\alpha} (\Gamma^{\mu \nu}\theta)_{\beta)}
(\overline{\theta}\Gamma_{\mu}\partial_{a} \theta)
-\frac{1}{12}(\Gamma_{\nu}\theta)_{(\alpha} (\Gamma_{\mu}\theta)_{\beta)}
(\overline{\theta}\Gamma^{\mu \nu}\partial_{a} \theta) 
\end{eqnarray} 

And the supersymmetry under which the above currents are invariant is
\begin{equation}
\delta \theta^{\alpha} = \epsilon^{\alpha},
\end{equation} 
\begin{equation} 
\delta  X^{\mu} = -\frac{1}{2} \overline{\epsilon} \Gamma_{\mu} \theta,
\end{equation} 
\begin{equation} 
\delta \phi_{\mu \nu} = -\frac{1}{2} \overline{\epsilon} 
\Gamma_{\mu \nu} \theta,
\end{equation}   
\begin{equation}
\delta \phi_{\mu \alpha} = -X^{\nu}(\Gamma_{\mu \nu} \epsilon)_{\alpha}
- \phi_{\mu \nu} (\Gamma^{\nu} \epsilon)_{\alpha} 
+ \frac{1}{6} (\overline{\epsilon} \Gamma_{\mu \nu} \theta) 
(\Gamma^{\nu} \theta)_{\alpha}
+ \frac{1}{6} (\overline{\epsilon} \Gamma^{\nu} \theta)
(\Gamma_{\mu \nu} \theta)_{\alpha},
\end{equation} 
\begin{eqnarray} 
\delta \phi_{\alpha \beta} = -\frac{1}{4} (\Gamma^{\mu})_{\alpha \beta}
\overline{\epsilon} \phi_{\mu} -
2(\Gamma^{\mu}\epsilon)_{(\alpha}\phi_{\mu \beta)}
-\frac{1}{4}X^{\mu}(\overline{\epsilon} \Gamma_{\mu \nu} \theta)
(\Gamma^{\nu})_{\alpha \beta}  \nonumber   \\   
-\frac{1}{4}X^{\mu}(\overline{\epsilon}\Gamma^{\nu}\theta)
(\Gamma_{\mu \nu})_{\alpha \beta} 
-\frac{1}{12}\overline{\epsilon}\Gamma_{\mu \nu}\theta 
(\Gamma^{\mu}\theta)_{(\alpha}(\Gamma^{\nu}\theta)_{\beta)} \nonumber \\   
-\frac{1}{12}\overline{\epsilon}\Gamma_{\mu}\theta 
(\Gamma^{\mu \nu}\theta)_{(\alpha}(\Gamma_{\nu}\theta)_{\beta)}     
\end{eqnarray} 

Such transformations are indeed a symmetry if the gamma matrices 
satisfy a condition which requires the dimensionality of the target
space to be 4, 5, 7 and 11.

Given those supersymmetric currents, Bergshoeff and Sezgin construct
the invariant action
\begin{equation}
S =T \int d^{3}x \sqrt{-\gamma} [ -\frac{1}{2}\gamma^{ab}L^{\mu}_{a} L_{b \mu}
+ \frac{1}{2}
- \frac{\varepsilon^{abc}}{\sqrt{-\gamma}}(L^{\mu}_{a}L^{\nu}_{b}L_{c \mu \nu}
+ \frac{9}{10}L^{\mu}_{a}L^{\alpha}_{b}L_{c \mu \alpha}
- \frac{1}{5}L^{\alpha}_{a}L^{\beta}_{b}L_{c \alpha \beta} )]  
\end{equation}  

The coefficients of the three last terms, which are cubic in the currents,
and which are contracted to $\varepsilon^{abc}$ are chosen so that all the 
dependence on the additional fields $\phi_{\mu \nu}$, $\phi_{\mu \alpha}$
and $\phi_{\alpha \beta}$ is through a total divergence. This is exactly
the total divergence by means of which we can once again define a composite
gauge field analogous to the one used in the bosonic case, as it was done
in the case of the superstring.

We now consider the $2+1$ brane action with a modified measure. For this we 
first get rid of the cosmological term and second consider the
change $T\sqrt{-\gamma}d^{3}x \rightarrow \Phi d^{3}x$, where 
\begin{equation}
\Phi \equiv \varepsilon^{abc}\varepsilon_{ijk} \partial_{a} \varphi_{i}
\partial_{b} \varphi_{j}\partial_{c} \varphi_{k} 
\end{equation}
That is, we consider the action,
\begin{equation}   
S =T \int d^{3}x \Phi [ -\frac{1}{2}\gamma^{ab}L^{\mu}_{a} L_{b \mu}
- \frac{\varepsilon^{abc}}{\sqrt{-\gamma}}(L^{\mu}_{a}L^{\nu}_{b}L_{c \mu \nu}                                                                  
+ \frac{9}{10}L^{\mu}_{a}L^{\alpha}_{b}L_{c \mu \alpha} 
- \frac{1}{5}L^{\alpha}_{a}L^{\beta}_{b}L_{c \alpha \beta} )] 
\end{equation}

In spite of the higher complexity, the basic structure of the theory and 
the way how the equations of motion work is that same as that of the 
superstring, explained in section 3.

As in any case, the variation with respect to the measure fields 
$\varphi_{i}$ imposes the constraint that the Lagrangian equals a 
constant, if $\Phi \neq 0$, that is 

\begin{eqnarray}
L = -\frac{1}{2}\gamma^{ab}L^{\mu}_{a} L_{b \mu}
- \frac{\varepsilon^{abc}}{\sqrt{-\gamma}}(L^{\mu}_{a}L^{\nu}_{b}L_{c \mu \nu} 
+ \frac{9}{10}L^{\mu}_{a}L^{\alpha}_{b}L_{c \mu \alpha}   \nonumber \\
- \frac{1}{5}L^{\alpha}_{a}L^{\beta}_{b}L_{c \alpha \beta} ) = M =constant
\end{eqnarray}                                                                               

Second, all the conditions obtained from extremizing with respect to 
variations of the fields $\phi_{\mu \nu}$, $\phi_{\mu \alpha}$ and 
$\phi_{\alpha \beta}$, are satisfied if 
\begin{equation}
\Phi = c\sqrt{-\gamma}
\end{equation}
where $c$ is a constant. From here we once again obtain that the brane 
tension appears as an integration constant.

The consideration of the equations obtained from the variation with 
respect to the world brane metric follow the same general structure to the
one discused in the bosonic case. 

Once this is realized, it is clear that, except for the existence of the 
constraint (67), all of the equations are the same as the ones we obtain in 
the Bergshoeff-Sezgin case \cite{Superb} after an appropriate rescaling
of the metric $\gamma_{ab}$, which is equivalent to making the choice
\begin{equation}
M = 1 - d = -1
\end{equation}  
(as discussed in section 2, $M$ is not invariant under scaling 
transformations and through the use of scalings it can be changed
continously).

The constraint (67) however is totally absent in the case of 
Ref. \cite{Superb}, where the fields
$\phi_{\mu \nu}$, $\phi_{\mu \alpha}$ and $\phi_{\alpha \beta}$, 
although playing an interesting group theoretical role are totally
irrelevant dynamically and therefore totally undetermined.

\section{The case of higher branes with a modified measure}
It is clear that for higher branes, once the Bergshoeff Sezgin 
construction is known \cite{Superb}, the two operations quoted in the 
case of the $2+1$ superbrane could also apply, that is: take the
Bergshoeff Sezgin lagrangian, first eliminate the cosmological term
and second, modify the integration measure (in a way that generalizes
straightforwardly from what we have done in the string and in the $2+1$ brane)
by making the replacement $T\sqrt{-\gamma}d^{d+1}x \rightarrow \Phi d^{d+1}x$,
with  $\Phi$ given as in eq. (23). 

Then the gauge fields, which we had to introduce in the bosonic case in 
order to have a consistent dynamics, are provided by the extra fields 
required by the Bergshoeff Sezgin formalism, who got to these constructions
from a group theoretic point of view \cite{Superb}.
                                                   
\section{Discussion and Conclusions}
In this paper, we have seen that a formulation of superstrings and 
superbranes with a modified measure is possible.

Due to the construction of this measure as $\Phi d^{d+1}x$, to the lagrangian 
that multiplies this structure we can add an arbitrary constant, since
$\Phi $ is a total derivative. In this sense, the origin of the vacuum
energy density need not be specified in the theory. It may appear through
the initial conditions.

In these theories, the tension of the string or brane appears as an 
integration constant.

Furthermore, such a formulation appears to give a dynamical role and not just
a group theoretical role to the extra fields introduced by Siegel 
\cite{Siegel} and Bergoshoeff and Sezgin \cite{Superb}.

This may be important in the quantization of the theory and may be also
important in the consequences for the low energy gravitational theory
that follows from these kind of brane theories. Recall that the original
motivation for introducing a modified measure was in this context 
\cite{modmes}.

Finally, a very interesting phenomena takes place in the formalism studied
here, which is the fact that what we used to think was a total divergence
becomes dynamically relevant, even at the classical level and beyond 
purely topological effects. This is of course due to the use of
the modified measure. Such observation raises new possibilities concerning
the study and resolution of fundamental questions concerning the dynamical
role of total divergences like in the strong CP problem. Some observations 
concerning a possible resolution of the strong CP problem by the use of
composite scalar field structures have been made already in the last paper of
Ref \cite{modmes}.
 
\section{Acknowledgements}  I  want to thank J.Bekenstein, R.Brustein, 
C.Castro, S. de Alwis, A.Davidson, A.Kaganovich, E.Nissimov, S.Pacheva, 
J.Portnoy  and L.C.R. Wijewardhana for discussions.


\begin{thebibliography}{99}

\bibitem{Wein}
For reviews see S.Weinberg, Rev. Mod. Phys.. 61, 1 (1989); 
Y.J.Ng., Int. J. Mod. Phys. D1, 145 (1992) 
and S.Carroll, astro-ph/0004075. 

\bibitem{modmes} For a review of theories of this type and further references
see, E.I.Guendelman and A.B.Kaganovich,
Phys. Rev. D 60, 065004 (1999) and for the connection of these theories 
to  theories of composite gauge fields and volume preserving diffeomorphisms 
developed in E.I.Guendelman, E.Nissimov and S.Pacheva, hep-th/9903245;  
E.I.Guendelman, E.Nissimov and S.Pacheva, Phys. Lett. B360, 57 (1995) and
C.Castro, Int. J. Mod. Phys. A13, 1263 (1998),  
see  E.I.Guendelman Int. Journ. Mod. Phys. A14, 3497 (1999).   


\bibitem{ind-mea} E.I.Guendelman, gr-qc/0004011. 

\bibitem{decl} E.I.Guendelman, 
Mod. Phys. Lett. A14, 1043 (1999).

\bibitem{cosco} E.I.Guendelman, 
Mod. Phys. Lett. A14, 1397 (1999);  E.I.Guendelman, gr-qc/9901067; 
E.I.Guendelman, "Scale Invariance and Cosmology", in Proc. of 
the 8-th Canadian Conference 
on General Relativity and Relativistic Astrophysics, Montreal, pp. 201 (1999).

\bibitem{gold} E.I.Guendelman, 
Class. Quantum Grav. 17, 361 (2000).

\bibitem{Yuval} For an introduction see Y.Ne'eman and E.Eizenberg,  
Membranes and Other Extendons, World Scientific, 1995. 

\bibitem{strm} E.I.Guendelman, STRINGS AND BRANES WITH A MODIFIED MEASURE.
e-Print Archive: hep-th/0005041. 

\bibitem{Poly} A.M.Polyakov, Phys. Lett.  B103, 207 (1981); L.Brink, 
P.DiVecchia and P.S.Howe, Phys. Lett. B65, 471 (1976); S.Deser and B.Zumino,
Phys. Lett. B65, 369 (1976).
\bibitem{lambda} P.S.Howe and R.W.Tucker, J.Phys. A10, L155 (1977); 
A.Sugamoto, Nucl. Phys. B215, 381 (1983); E.Bergshoeff, E.Sezgin and
P.K.Townsend, Phys. Lett. 189B, 75 (1987); A.Achucarro. J.M.Evans, 
P.K.Townsend and D.L.Wiltshire, Phys. Lett. 198B, 441 (1987).  
\bibitem{Dolan} B.P.Dolan and D.H.Tchrakian, Phys. Lett. B202, 211 (1988).
\bibitem{Barc} M.S.Alves and J.Barcelos-Neto, Europh. lett. 7, 395 (1988);
J.Barcelos-Neto, Phys. Lett. 245B, 26 (1990) and Phys. Lett. 249B, 551 (1990);
U. Lindstrom  and G.Theodoridis , Phys. Lett. 208B, 407 (1988);
A.Kalhede  and U.Lindstrom, Phys. Lett. 209B, 441 (1988).
\bibitem{Eiz} E.Eizenberg and Y.Ne'eman, Il Nuovo Cimento, 102, 1183 (1989).
\bibitem{Berg} E.Bergshoeff, L.A.London and P.K.Townsend, Class. Quantum 
Grav. 9, 2545 (1992).
\bibitem{Anso} S.Ansoldi, C.Castro and E.Spallucci, hep-th/0005132.
\bibitem{GS} M.B.Green and J.H.Shwarz, Phys. Lett. 136B, 367 (1984); 
Nucl. Phys. B243, 285 (1984). 
\bibitem{Siegel} W.Siegel, Phys. Rev. D50, 2799 (1994).
\bibitem{Superb} E.Bergshoeff and E. Sezgin, Phys. Lett. 354B, 256 (1995).
\end{thebibliography}
\end{document}